\documentclass[aps, twocolumn, pra, showpacs, amsmath, superscriptaddress]{revtex4}
\usepackage{graphicx}
\usepackage{dcolumn}
\usepackage{bm}
\usepackage{amsmath}
\usepackage{subfigure}
\usepackage{amsfonts}
\usepackage{xcolor}
\usepackage{appendix}
\usepackage{multirow}

\begin{document}

\title{Atomic-scale Positioning of Single Spins via Multiple Nitrogen-vacancy Centers}
\author{Wen-Long Ma}
\affiliation{State Key Laboratory of Superlattices and Microstructures, Institute of Semiconductors, Chinese Academy of Sciences, Beijing, 100083, China}
\affiliation{Department of Physics, The Chinese University of Hong Kong, Shatin, N. T., Hong Kong, China}
\author{Shu-Shen Li}
\affiliation{State Key Laboratory of Superlattices and Microstructures, Institute of Semiconductors, Chinese Academy of Sciences, Beijing, 100083, China}
\author{Geng-Yu Cao}
\affiliation{State Key Laboratory of Magnetic Resonance and Atomic and Molecular Physics, Wuhan Institute of physics and Mathematics, Chinese Academy of Sciences, Wuhan, 430071, China}
\author{Ren-Bao Liu}
\affiliation{Department of Physics, The Chinese University of Hong Kong, Shatin, N. T., Hong Kong, China}
\affiliation{Centre for Quantum Coherence, The Chinese University of Hong Kong, Shatin, N.T., Hong Kong, China}
\affiliation{Institute of Theoretical Physics, The Chinese University of Hong Kong, Shatin, N.T., Hong Kong, China}
\email{rbliu@phys.cuhk.edu.hk}

\date{\today }

\begin{abstract}
 We present a scheme of positioning a single electron spin with sub-nanometer resolution through multiple nitrogen-vacancy centers in diamond. With unwanted noise suppressed by dynamical decoupling, the spin coherence of each center develops characteristic oscillations due to a single electron spin located $4{\sim20}$ nm away from the centers. We can extract the position information from the characteristic electron spin-coherence oscillations of each center. This scheme is useful for high-resolution nanoscale magnetometry.

\end{abstract}
\maketitle

\section{Introduction}
Magnetic resonance spectroscopy of ensembles of spins has wide applications in many fields such as material science, analytical chemistry, structural biology \cite{book-1}. Pushing the sensitivity of conventional magnetic resonance imaging to the single spin level would enable even more important application in single molecule structure analysis \cite{molecule-1,nsb}. However, it is extremely difficult to detect the weak magnetic field from a single electron or nuclear spin by conventional magnetic resonance spectroscopy \cite{n1,npa}. The magnetic resonance force spectroscopy has been developed to improve the sensitivity to a few spins but with requirements of low temperature and high vacuum \cite{apl1,pnas-1}, which limit its application to biological systems.


\begin{figure*}
\includegraphics[width=2.5in]{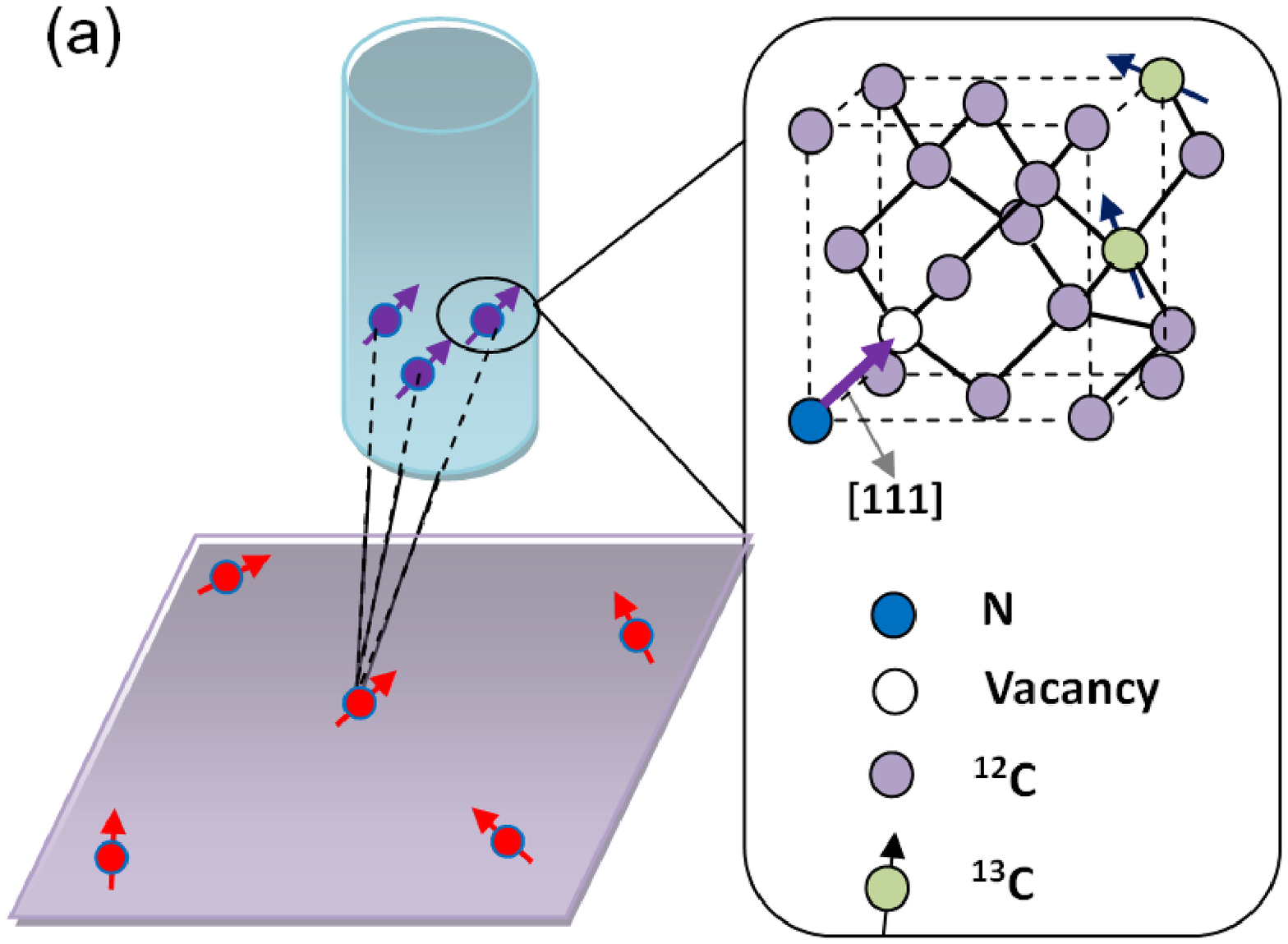}
\includegraphics[width=1.9in]{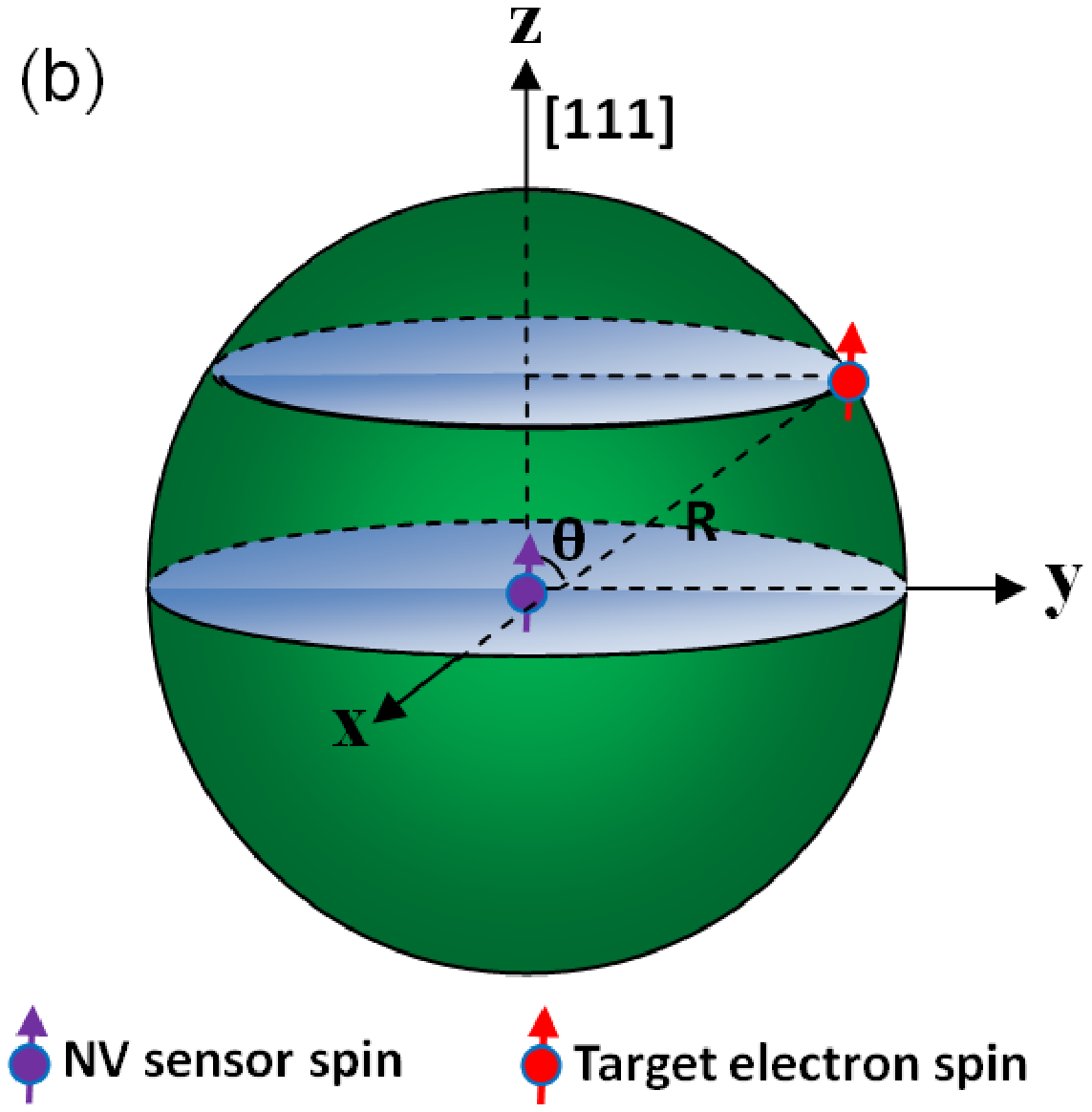}
\includegraphics[width=2.7in]{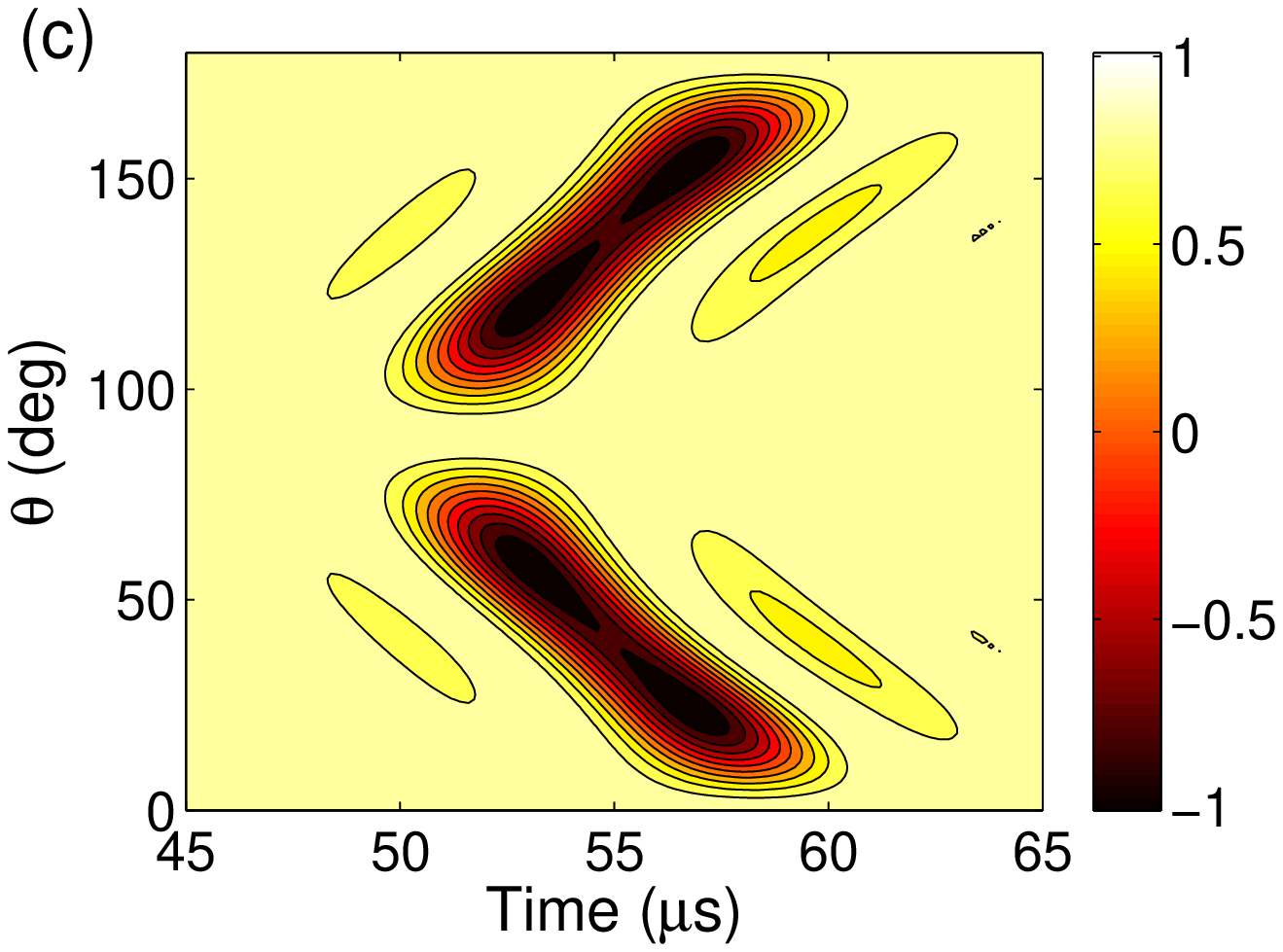}
\includegraphics[width=2.7in]{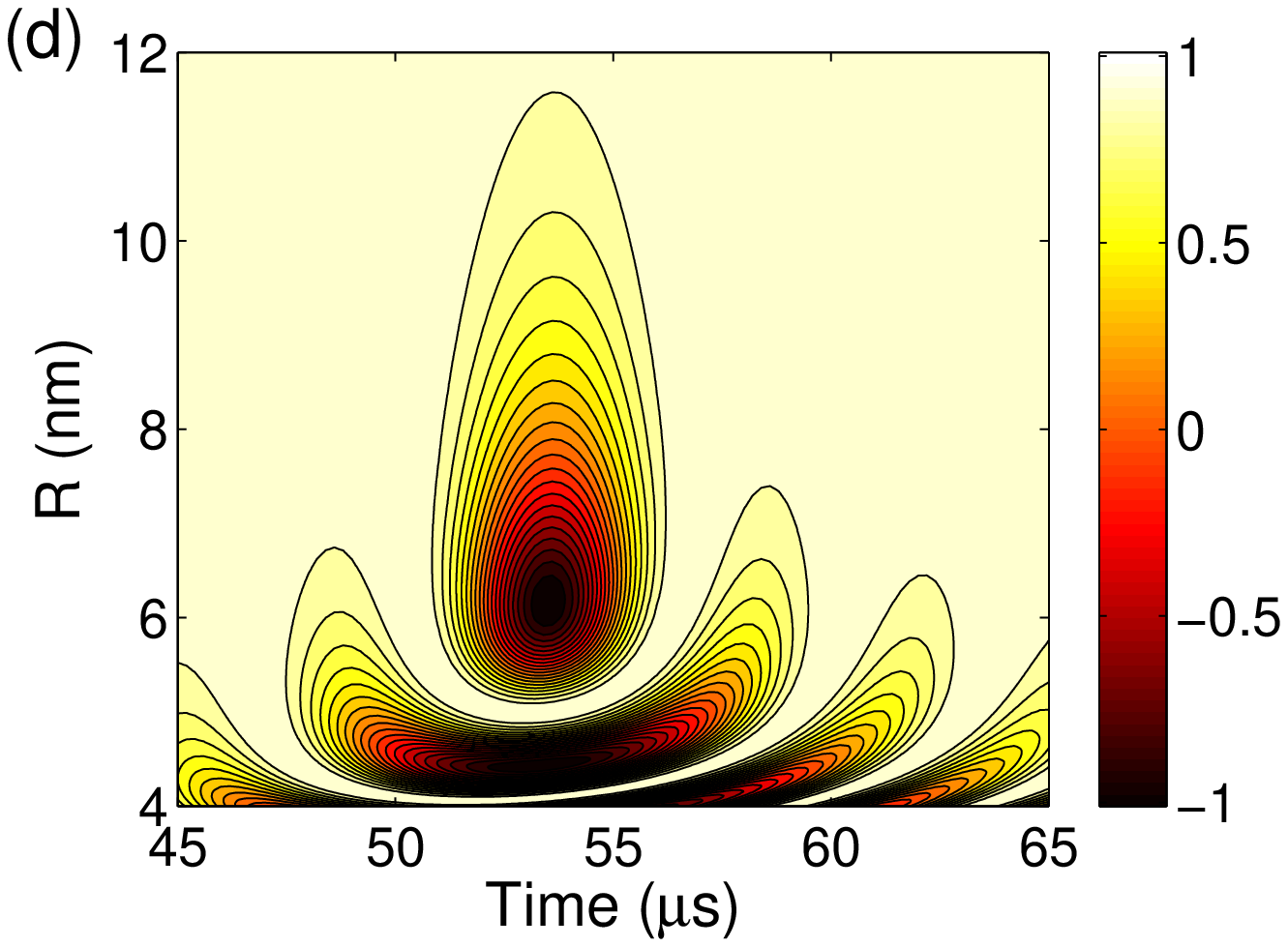}
\centering
\caption{Fingerprint features in NV center spin coherence due to coupling to a single electron spin. (a) Schematic illustration of the single electron spin detection based on three NV centers embedded in a diamond tip. The NV axes of the three NV centers are all chosen along the crystal axis [111]. The close-up shows the atomic structure of the NV center. (b) The relative position between a single NV sensor spin and the target electron spin is denoted by the distance $R$ and the zenith angle $\theta$ relative to the [111] axis. (c) NV center spin coherence under CPMG-30 control as a function of time and the zenith angle of the target spin at a distance of $R=5$ nm. (d) NV center spin coherence under CPMG-30 control as a function of time and the distance $R$ of the target spin with a zenith angle $\theta=30^{\circ}$. The strain-induced transverse anisotropy parameters are $\varepsilon_{A}=3$ MHz, $\varepsilon_{B}=2$ MHz, $\varepsilon_{C}=4$ MHz and a magnetic field $B=0.1$ Gauss is applied along the NV axis. The target electron spin is assumed to have the same gyromagnetic ratio as the NV electron spins ($\gamma_{e}=\gamma_{\text{NV}}$). The calculations in (c) and (d) are performed for one of the three NV centers (NV-A) with only the coupling between NV-A and the target spin considered. }
\label{contour}
\end{figure*}

Recently single spin detection in ambient conditions has been made possible by employing the atomic-scale defect in diamond---the nitrogen-vacancy (NV) center \cite{prl1,np2,s1,n2}---as a magnetic sensor. The NV center has a paramagnetic spin-triplet ground state that can be optically initialized and read out \cite{prb1,prl2} and has a long spin coherence time up to milliseconds even at room temperature \cite{nm1}. With recent development in diamond nanofabrication \cite{nano1}, the NV center in high-purity diamond has been used as a magnetic sensor to detect single nuclear spins \cite{nano2,prlb,prlc} and nuclear spin clusters \cite{cluster} in diamond. The shallow NV centers below the diamond surface have also been used to sense single electron spins in single molecules \cite{electron-1,electron-2} and electron spin labels in single proteins \cite{protein-1} on the surface. The high-resolution positioning of single spins has potential applications in revealing the structure and dynamics of single molecules \cite{ar1}.

A widely-used sensing scheme is the quantum sensing based on dynamical decoupling (DD) \cite{DD-1,DD-2} based quantum sensing. The key idea is to identify the characteristic oscillations caused by the target spin imprinted onto electron spin coherence of the NV center when the noise from the target spin is resonantly amplified by DD control of the NV electron spin \cite{DDsensing-1,nano3}. The DD control can also suppress the background noise and prolong the coherence time of NV electron spin. However, the position of the target spin cannot be uniquely determined by one experimental setting \cite{nano2,nano3}. Instead, the measurement has to be repeated for various magnetic field orientations or for different scanning positions of the sensor to fully position the target spin. Such a requirement restricts the speed of scanning and limits the potential applications in single-molecule structure imaging.

In this paper, we propose to use multiple NV centers in a diamond probe tip for positioning single spins with sub-nanometer resolution [Fig. \ref{contour}(a)]. This scheme is analogous to the multi-satellite positioning systems. For a given NV center spin, the coherent oscillations provide partial information on the distance and direction of the target single spin. The full position information of the target spin can be obtained by the oscillations of the multiple NV center spins. This scheme would greatly improve the speed of detection compared with the scheme using one NV center \cite{nano2,nano3,prlb,prlc}, since there is no need of varying the magnetic field or nanometer-step scanning the probe \cite{npa,nano1,ar1}.

\section{Theoretical model}
As shown in Fig. \ref{contour}(a), the proposed setup to detect a remote single electron spin contains three NV centers in a diamond dip. The spin Hamiltonian of the whole system is
\begin{align}\label{eq1}
   H=&H_{\rm{tar}}+\sum_{i}H_{\text{NV}_{i}}+H_{\rm{bath}}+\sum_{i,j}H_{\text{NV}_{i}-\text{NV}_{j}} \nonumber \\
     &+\sum_{i}H_{\text{NV}_{i}-\rm{tar}}+\sum_{i}H_{\text{NV}_{i}-\rm{bath}},
\end{align}
where the first three terms denote the Hamiltonians of the isolated target electron, the NV center sensors and the nuclear spin bath correspondingly, and the last three terms represent the interaction between different NV center sensors, the coupling of the NV center sensors to the target electron spin, and the coupling of the NV center sensors to the nuclear spin bath in diamond. We assume that the Hamiltonian of the spin-1/2 target electron spin is $H_{\rm{tar}}=-\gamma_{e}\mathbf{B}\cdot{\mathbf{S}_{e}}$ with $\gamma_{e}$ and ${\mathbf{S}_{e}}$ denoting the gyromagnetic ratio and spin operator of the target spin respectively, and the Hamiltonian of the $i$th NV center is \cite{npb,cpl}
\begin{eqnarray}\label{H-NV}
   H_{\text{NV}_{i}}=\Delta({S_{i}^{z}})^{2}+{\varepsilon}_{i}[(S_{i}^{x})^{2}-(S_{i}^{y})^{2}]-\gamma_{\text{NV}}\mathbf{B}\cdot{\mathbf{S}_{i}},
\end{eqnarray}
where ${\mathbf{S}_{i}}$ is electron spin operator of the $i$th NV center, $\Delta=2.87$ GHz is the zero-field splitting of the NV centers, $\varepsilon_{i}$ denotes the local strain-induced transverse anisotropy for the $i$th NV center, $\gamma_{\text{NV}}$=-2.8 MHz Gauss$^{-1}$ is the gyromagnetic ratio of NV electron spin, $\mathbf{B}$ is the magnetic field, and the $z$ axis is the intrinsic electron spin quantization axis of the $i$th NV center (pointing from the nitrogen to the vacancy). In this paper,

With a rather strong magnetic field applied along $z$ axis, the local strain terms in Eq. (\ref{H-NV}) can be neglected and the $i$th NV center has three spin eigenstates $\{|0\rangle_i, |\pm1\rangle_i\}$. However, in the weak magnetic field regime, the strain-induced transverse anisotropy breaks the degeneracy of the $|\pm1\rangle_i$ states in zero magnetic field (the eigenstate $|0\rangle_i$ is almost unperturbed since $\Delta_{i}\gg\varepsilon_{i}$) and results in renormalized eigenstates $|\pm\rangle_i$ \cite{SI,prbr3} (see Appendix A). In the new eigenstate basis $\{|0\rangle_i, |+\rangle_i, |-\rangle_i\}$ of the $i$th NV center, the full Hamiltonian in Eq. (1) can be recast into the pure-dephasing form as (see Appendix B)
\begin{eqnarray}\label{HR}
   H_i=H_i^{(+)}|+\rangle_i\langle+|+H_i^{(-)}|-\rangle_i\langle-|+H_i^{(0)}|0\rangle_i\langle0|,
\end{eqnarray}
with
\begin{align}\label{}
H_{i}^{(\pm)}\approx&H_i^{(0)}\pm\left(\sqrt{\varepsilon_{i}^{2}+\omega_\mathrm{NV}^{2}}+\frac{\omega_\mathrm{NV} h_i}{\sqrt{\varepsilon_{i}^{2}+\omega_\mathrm{NV}^{2}}}
+\frac{h_i^{2}}{2\sqrt{\varepsilon_{i}^{2}+\omega_\mathrm{NV}^{2}}}\right), \label{HRpm}\\
H_i^{(0)}=&H_{\text{tar}}+\sum_{j,k\neq{i}}H_{\mathrm{NV}_{j}-\mathrm{NV}_{k}}+H_{\text{bath}} \nonumber \\
&+\sum_{j\neq{i}}(H_{\mathrm{NV}_{j}}+H_{\mathrm{NV}_{j}-\text{bath}}),
\end{align}
where $\omega_{\mathrm{NV}}=|\gamma_{\mathrm{NV}}|B$ is the Larmor frequency of the NV electron spin, $h_i=h_i^{e}+h_i^{b}+h_i^{s}$ is the noise field for the $i$th NV center,  $h_i^{e}=(\hat{\mathbf{z}}\cdot{\mathbb{A}_{i,e}})\cdot{\mathbf{S}_{e}}$ is the dipolar magnetic field produced by the target electron spin, $h_i^{b}=\sum_{m}(\hat{\mathbf{z}}\cdot{\mathbb{A}_{i,m}})\cdot{\mathbf{I}_{m}}$ is the nuclear Overhauster field with ${\mathbf{I}_{m}}$ denoting the nuclear spin operator of the $m$th nuclear spin in the bath, $h_i^{s}=\sum_{j\neq{i}}(\hat{\mathbf{z}}\cdot{\mathbb{A}_{i,j}})\cdot{\mathbf{S}_{j}}$ is the dipolar magnetic field produced by the other NV centers, $\mathbb{A}_{i,e},\mathbb{A}_{i,m},\mathbb{A}_{i,j}$ are the dipolar interaction tensors for the coupling of the $i$th NV electron spin to the target spin, the $m$th nuclear spin and the $j$th NV electron spin correspondingly, and $H_{\text{bath}}=-\gamma_{n}\mathbf{B}\cdot\sum_{m}{\mathbf{{I}}_{m}}+\sum_{m<n}\mathbf{I}_{m}\cdot\mathbb{D}_{mn}\cdot\mathbf{I}_{n}$ where $\gamma_{n}$ is the gyromangetic ratio of nuclear spins and ${\mathbb{D}_{mn}}$ is the nuclear-nuclear dipolar interaction tensor.

As shown in Eq. (\ref{HRpm}), the renormalized eigenstates $|\pm\rangle_{i}$ have eigenenergies $\pm\sqrt{\varepsilon^{2}_{i}+\omega_\mathrm{NV}^{2}}$. In the renormalized eigenstates $|\pm\rangle_{i}$, the coupling strength of the $i$th NV center to the target spin, the other NV centers and the nuclear spin bath are all reduced to a factor of $\omega_\mathrm{NV}/\sqrt{\varepsilon^{2}_{i}+\omega_\mathrm{NV}^{2}}$. Moreover, The last term in Eq. (4) mainly induces long-range interaction within the nuclear spin bath, but its effect on the sensor coherence can be largely suppressed for not too small magnetic field and large strain parameters. In this paper we consider the single transition ($|0\rangle\leftrightarrow|+\rangle_{i}$) of all the NV sensors.

The NV center spin decoherence in high-purity diamond is mainly caused by the hyperfine interaction with the $^{13}$C nuclear spins \cite{prb2,prl3}. At finite temperature, the random orientations of the nuclear spins result in local field fluctuation (thermal fluctuation) that can be eliminated by spin echo. The internal dynamics of the spin bath also induces the dynamical quantum fluctuation \cite{prb2}, which cannot be completely removed by dynamical decoupling. To detect a single electron spin by the NV center, we should suppress the thermal and quantum noise caused by the nuclear spin bath while amplifying the effect of the target electron on the NV center spin coherence. The Carr-Purcell-Meiboom-Gill (CPMG) control \cite{nano3,s2,prl4} meets both requirements, since it can simultaneously increase the coherence time of the sensor spin while selectively amplifying the noise at a specific frequency. For the $N$-pulse CPMG (CPMG-$N$) sequence, the NV electron spin is flipped at time $t_{k}=(2k-1)/(2N)$ with $t$ being the total evolution time and $k=1,2,\cdots,{N}$. The target spin can be detected if the NV spin coherence profile develops characteristic oscillations caused by the target spin.

 The oscillation features in the NV sensor spin coherence caused by the target spin can be understood as follows.
 In the classical noise picture, the precession of the single electron spin near the NV sensor gives rise to additional peak structures \cite{nano3} in the smooth noise spectrum of the nuclear spin bath \cite{prb3}, corresponding to a series of coherence dips in the NV sensor spin decoherence profile. When the magnetic field is much stronger than the dipolar interaction with the target spin, the peak in the noise spectrum is approximately located at the electron spin Larmor frequency $\omega_e=|\gamma_{e}|B$. For CPMG-$N$ control, the coherence dips of the $i$th NV sensor spin caused by a remote target electron spin occur approximately at times
 \begin{align}\label{}
  t^i_{\text{dip}}=\frac{\pi(2q-1)N}{\omega_e+\lambda_i A_{i,e}^{z}/2},
  \label{tdip}
 \end{align}
 where $q=1,2,\cdots$ denotes the dip order. In the following we will always consider the first-order coherence dip ($q=1$). The sensor coherence dip depth as a function of the CPMG pulse number can be derived in the weak sensor-target coupling regime as \cite{dip}
 \begin{align}\label{}
 L^i_{\text{dip}}(N){\approx}\cos\left(\frac{\lambda_iA_{i,e}^{\bot}N}{\omega_e+\lambda_i A_{i,e}^{z}/2}\right),
 \label{cdip}
 \end{align}
 where $\lambda_i=\omega_\mathrm{NV}/\sqrt{\varepsilon^{2}_{i}+\omega_\mathrm{NV}^{2}}$ is the renormalization factor for the $i$th NV sensor, $A_{i,e}^{z}$, $A_{i,e}^{\perp}$ are the components of the dipolar interaction with the target spin parallel and orthogonal to the $z$ axis respectively. For a small pulse number, the above equation agrees with that from the semiclassical noise model $L^i_{\text{dip}}(N){\approx}\exp[-N^{2}(\lambda_iA_{i,e}^{\bot})^{2}/2(\omega_e+\lambda_i A_{i,e}^{z}/2)^2]$ (see Appendix C and D for details).

\section{Detection scheme}

\begin{figure}
\begin{minipage}[t]{0.5\linewidth}
\centering
\includegraphics[width=1.8in]{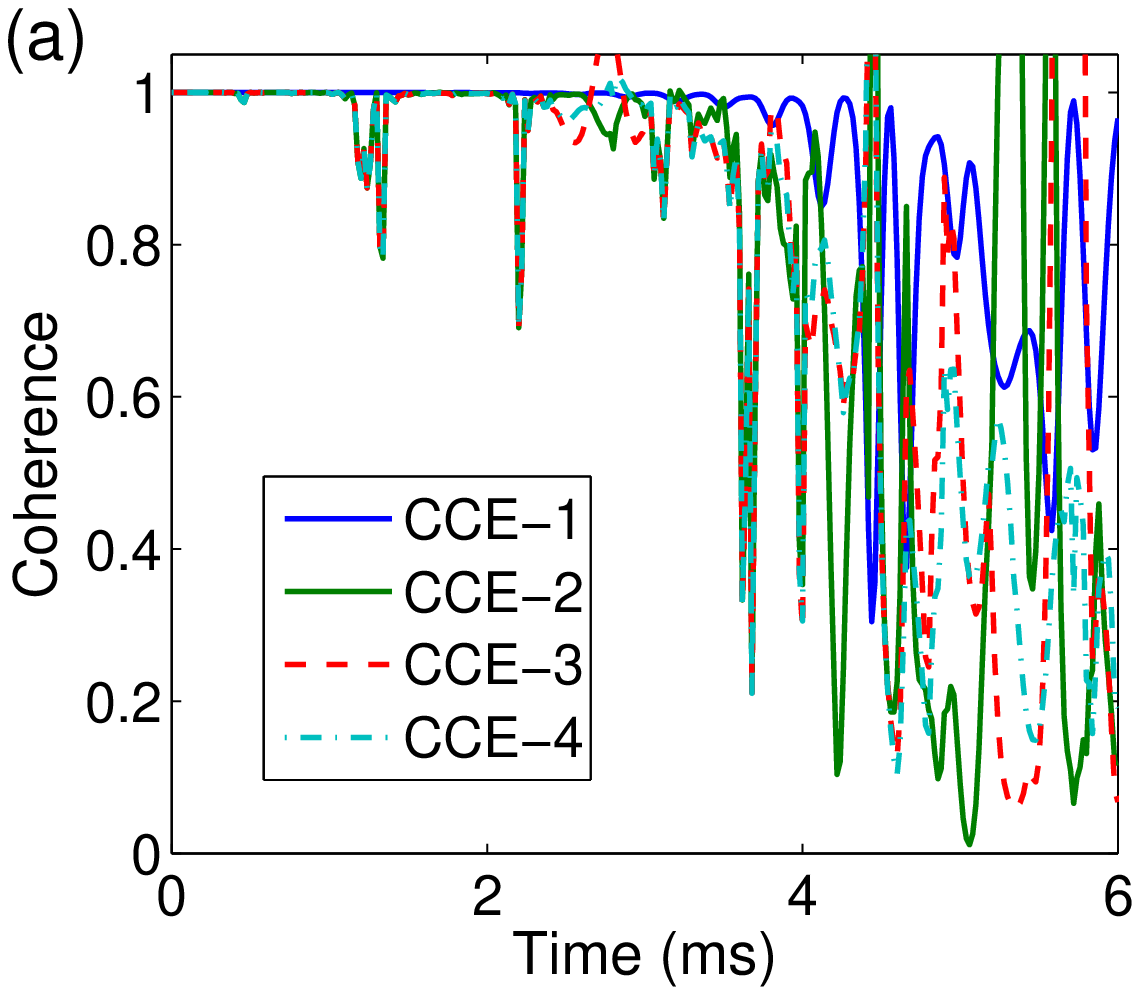}
\end{minipage}%
\begin{minipage}[t]{0.5\linewidth}
\includegraphics[width=1.8in]{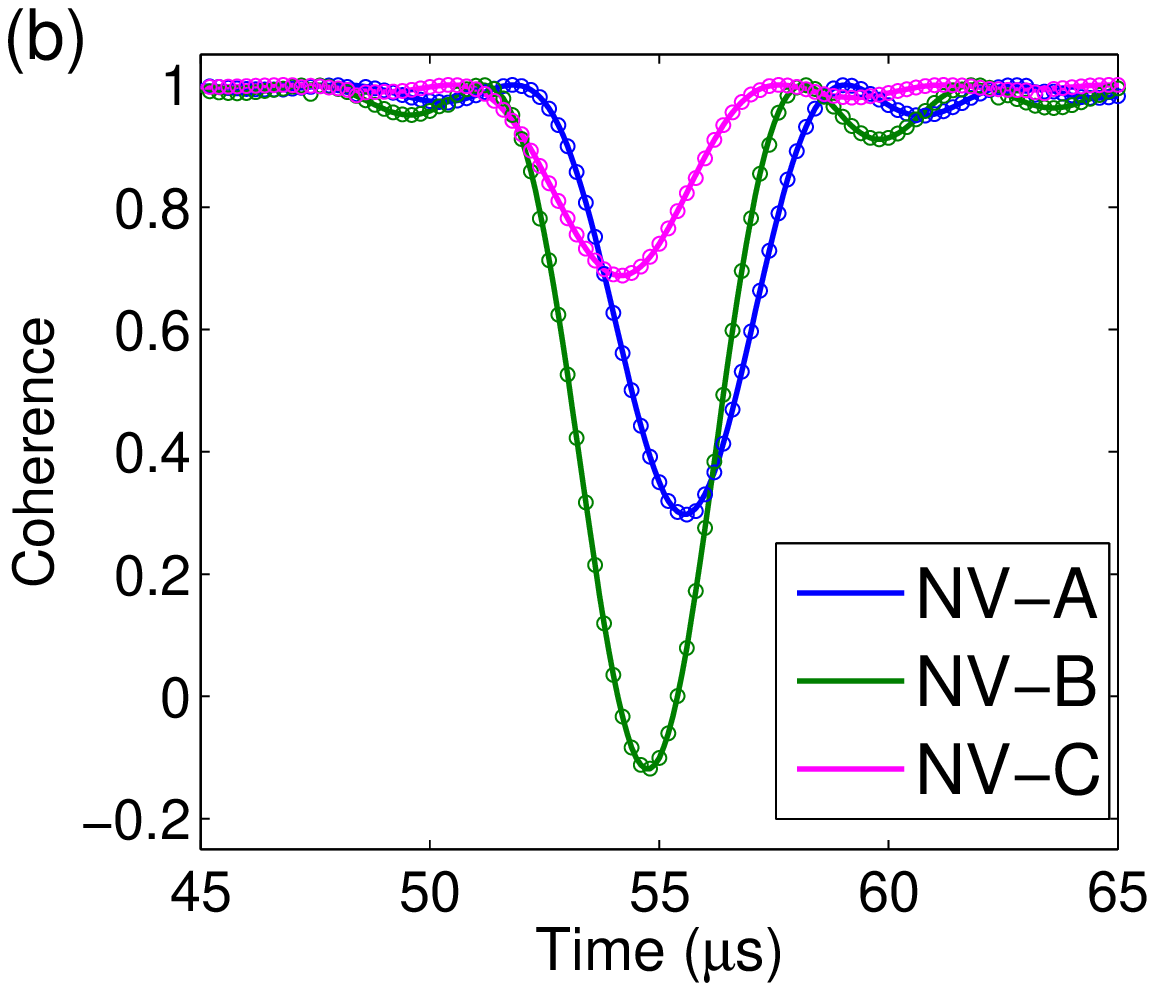}
\end{minipage}
\centering
\caption{Oscillations caused by the target spin on the NV center spin decoherence with the nuclear spin noise suppressed by dynamical decoupling. (a) Decoherence of NV-A center (in the absence of the target spin and other NV centers) caused by the $^{13}$C spin bath with a natural abundance (1.1\%) under CPMG-30 control. (b) Spin coherence of three NV centers (with the presence of the target spin) under CPMG-30 control (circles) which matches the contribution solely from the target electron (solid lines) located at $(R_{A}=7.46$ {\text{nm}}, $ \theta_{A}=19.56^{\circ}, R_{B}=8.72$ {\text{nm}}, $\theta_{B}=33.92^{\circ}$ and $R_{C}=8.83$ {\text{nm}}, $ \theta_{C}=35.03^{\circ})$. The parameters are the same as those in Fig. \ref{contour}. }
\label{three}
\end{figure}

 To detect the target spin with high spatial resolution, the target spins in different positions should have distinguishable fingerprint oscillations imprinted onto the sensor spin coherence. This implies that the applied magnetic field should be comparable to the dipolar interaction between the sensor spin and the target spin. For detecting a single electron spin 5$\sim$10 nm away, the magnetic field should be about 0.2$\sim$0.05 Gauss. The renormalization factor is about 0.18$\sim$0.05 (the strain-induced transverse anisotropy \cite{s3} $\sim$ 3 MHz), which means the effective dipolar interaction is about ten times smaller than the magnetic field. To increase the detection resolution, we should use the CPMG control with a large number of pulses.

The renormalized dipolar interaction between the $i$th sensor spin and the target spin has components $[A_{R}^{z}, A_{R}^{\perp}]=\frac{\mu_{0}\gamma_{\mathrm{NV}}\gamma_{e}\lambda_i}{4{\pi}R^{3}}[1-3{\cos}^{2}\theta, 3\sin\theta\cos\theta]$ (where $\mu_{0}$ is the vacuum permeability) that are determined by the distance $R$ and the zenith angle $\theta$ between the displacement and the [111] axis [Fig. \ref{contour}(b)]. Figure \ref{contour}(c) and \ref{contour}(d) show the different coherence oscillations, from which we can see that an atomic-scale change of the position of the target spin would have noticeable influence on the time-domain coherence features of the NV sensor spin. It is worth noting that the contour plot in Fig. \ref{contour}(c) is mirror symmetric about the $\theta=90^{\circ}$ line owing to the symmetry of the dipolar interaction. We may use the coherence features to determine the distance and zenith angle of the target spin.

To fully determine the three-dimensional position of the target spin, we propose to use multiple NV centers as the sensor spins, in analogue to the multi-satellite global positioning systems. We note that recently strongly coupled NV pairs below the diamond suface with the coupling strength up to 53 kHz (corresponding to the distance about ~10 nm) and an average depth of 15 nm have been successfully generated by implanting ionized nitrogen molecules ($^{15}$N${^{+}_2}$) into diamond \cite{YamamotoPRB2013}. In principle, multiple closely spaced NV centers can be generated by implanting N$_3$ and N$_4$ molecules into diamond \cite{YamamotoPRB2013}.

We consider three NV sensor spins (A, B, C) embedded near the surface of a diamond tip. The three NV sensors can be separately addressed by their different energy splittings between $|0\rangle_i$ and $|+\rangle_i$ due to the difference in the local strain that usually occurs in real diamond \cite{s3}. In the simulation, we assume the strain-induced transverse anisotropy parameters of the three NV sensors to be $\varepsilon_{A}=3$ MHz, $\varepsilon_{B}=2$ MHz and $\varepsilon_{C}=4$ MHz. The differences in the local strain also help suppress the exchange interaction between the NV sensor spins in the multi-spin sensor, which simplifies the data analysis for the positioning. We may use the CPMG control with a specific frequency to induce the spin transition in one NV sensor while leaving the states of the other NV sensors unchanged. We get three sets of coordinates $(R_{A}, \theta_{A}), (R_{B}, \theta_{B}), (R_{C}, \theta_{C})$ by comparing the coherent oscillations due to the target spin with the dip features in the spin decoherence of the three NV sensors. Therefore the three-dimensional coordinates of the target spin are determined.

For simplicity, we suppose the three NV sensors are all along the crystal axis [111]. If they have different directions in the diamond lattice, the three sets of coordinates obtained by the detection scheme are with respect to the three different local coordinates of the three sensors, and need to be transformed into the same coordinate for positioning the target spin, and the effective Zeeman splittings relevant to the renormalization factors $\lambda_i$ of the NV sensors (projections of the magnetic field to the NV sensors' axes) are different for different sensors.

The NV sensors are separated from each other by $6\sim7$ nm with the dipolar interaction between them less than 0.1 MHz, while the energy cost of pairwise flip-flop processes (e.g. $|0\rangle_i |+\rangle_j \leftrightarrow |+\rangle_i |0\rangle_j$) between two NV sensors is approximated as $|\varepsilon_i-\varepsilon_j|\sim$ MHz, which is much stronger than the dipolar interaction between them. As a result, the flip-flop processes between different NV electron spins are highly suppressed \cite{SI}. Moreover, the single spin dynamics of the other NV sensors on the decoherence of a specific NV sensor is very small due to the much larger noise frequency from the $i$th NV sensor (corresponding to the transition $|+\rangle_i\leftrightarrow |-\rangle_i$) compared with target spin Larmor frequency \cite{SI}. Therefore for a specific NV sensor, the effects of other sensors on the sensor decoherence can be neglected.

\begin{figure}
\includegraphics[width=3.0in]{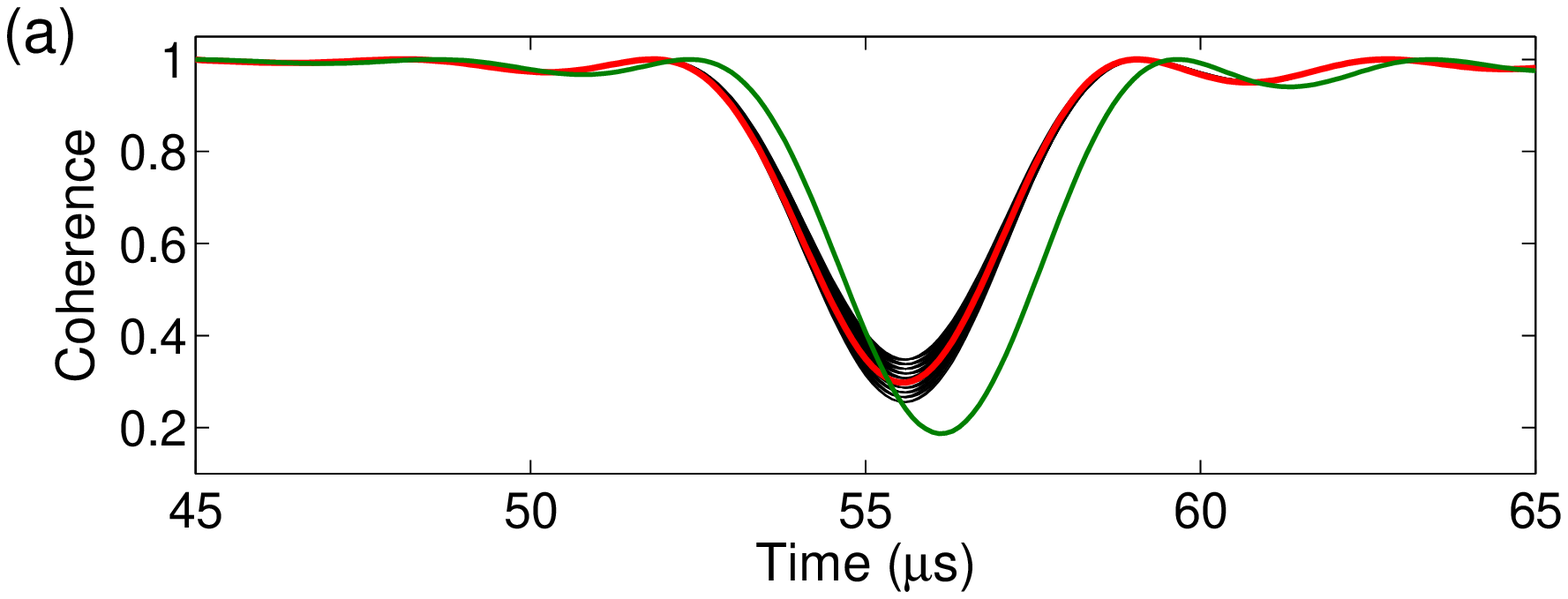}
\includegraphics[width=3.0in]{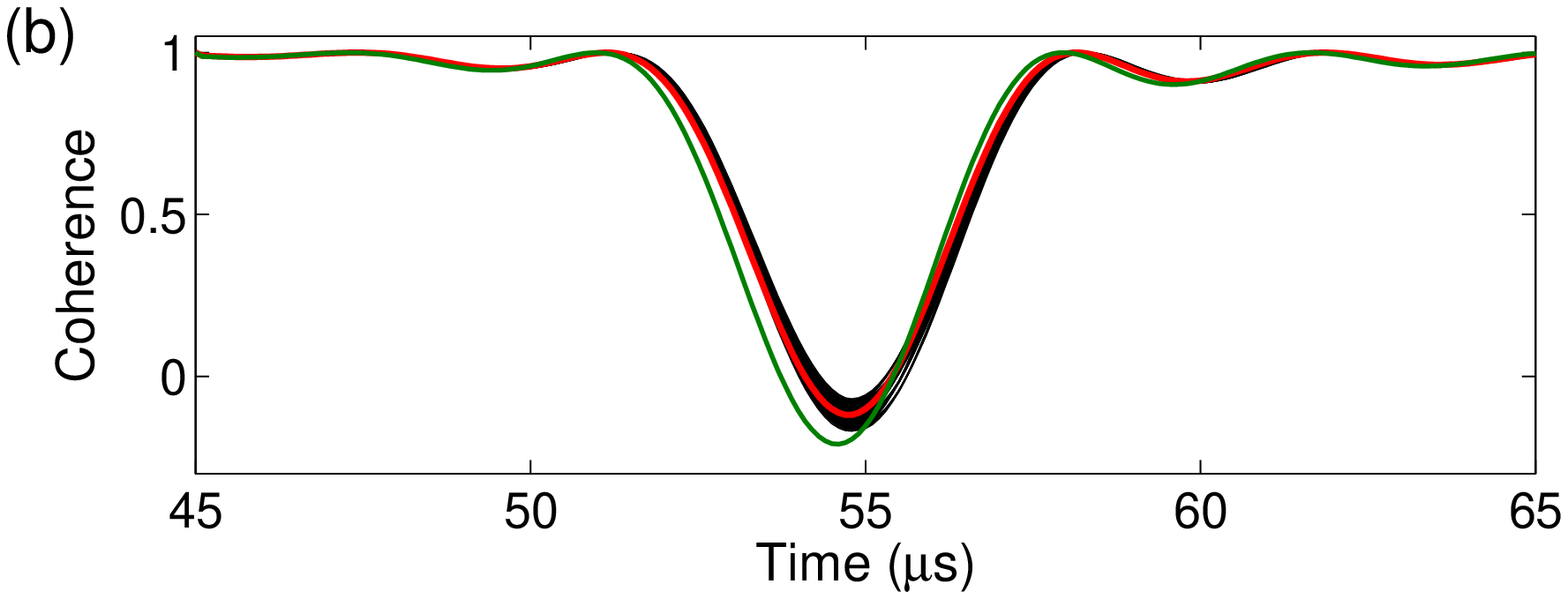}
\includegraphics[width=3.0in]{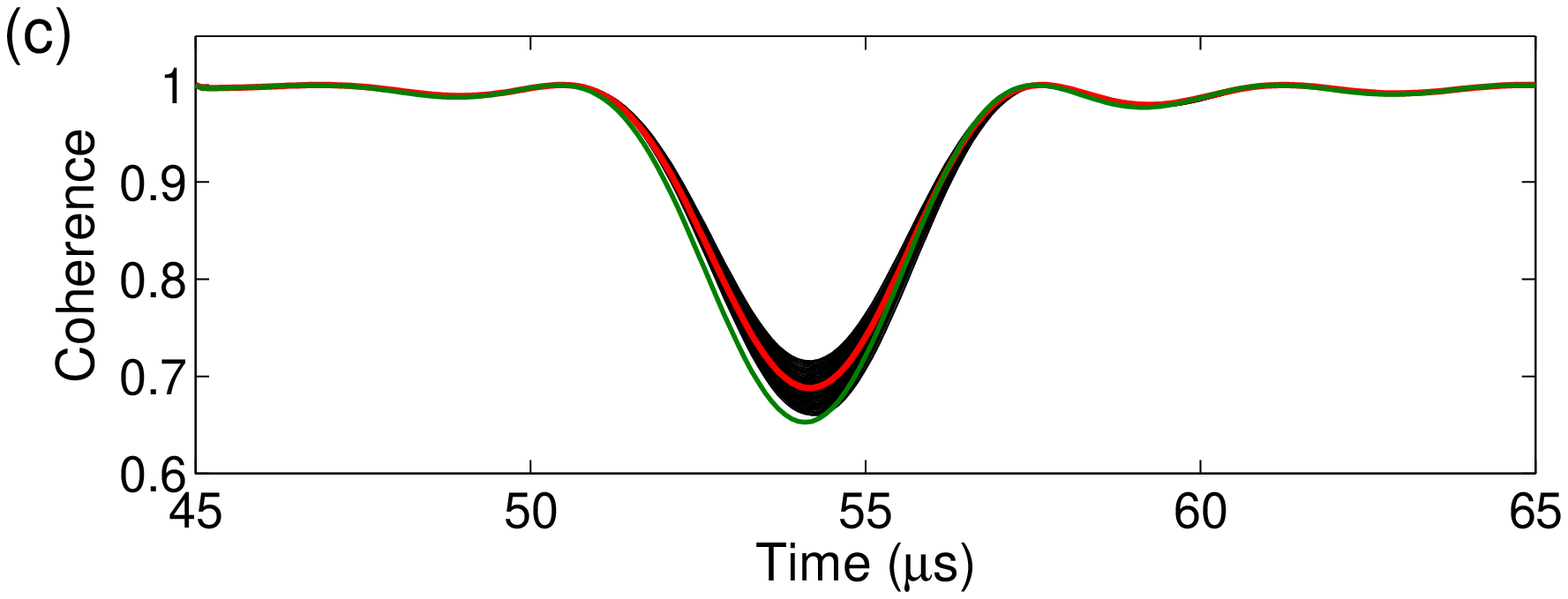}
\centering
\caption{Numerical simulation for positioning a target spin. The red lines represent the spin coherence obtained from the detection while the black solid lines represent the matched oscillation patterns in the fingerprint library (the discrete steps are taken as $dR=0.02$ {\text{nm}}, $d\theta=0.2^{\circ}$ here). The position and depth of the first dip in the spin coherence oscillations are used as the criteria for matching. The estimated locations of the electron relative to the three NV centers are (a) $R_{A}\in[7.40\sim7.50$ {\text{nm}}], $\theta_{A}\in[18.4^{\circ}\sim20.0^{\circ}]$, (b) $R_{B}\in[8.64\sim8.76$ {\text{nm}}], $\theta_{B}\in[32.2^{\circ}\sim34.2^{\circ}]$  and (c) $R_{C}\in[8.64\sim8.96$ {\text{nm}}], $\theta_{C}\in[33.0^{\circ}\sim36.0^{\circ}]$. The exact position of the target spin is such that $(R_{A}=7.46$ {\text{nm}}, $ \theta_{A}=19.56^{\circ}, R_{B}=8.72$ {\text{nm}}, $\theta_{B}=33.92^{\circ}$ and $R_{C}=8.83$ {\text{nm}}, $ \theta_{C}=35.03^{\circ})$. The green lines represent the oscillation patterns if the target spin is moved about 0.6 nm away from its original position. The parameters are the same as those in Fig. \ref{contour}
.}
\label{matching}
\end{figure}

We solve the pure dephasing problem with inclusion of the interactions between the NV sensors, the coupling to the target electron spin, and the coupling to the nuclear spin bath, by adopting the well-established cluster-correlation expansion method (CCE)\cite{prb2,re1,prb4,SI}. The key idea is that the qubit coherence can be expressed as the product of cluster correlations. In real calculations, it often suffices to truncate the expansion up to the minimum size $M$ of the clusters (CCE-$M$) to get converged results. It is clear from Fig. \ref{three}(a) that the spin coherence of a single NV sensor in $^{13}$C nuclear spin bath with a natural abundance 1.1\% can be well protected for $t<1$ ms under CPMG-30 control.

Apart from the $^{13}$C nuclear spin bath, the electron spin bath on the diamond surface cause additional decoherence for the near-surface NV centers \cite{RomachPRL2015,MyersPRL2014,RosskopfPRL2014,SushkovPRL2014}. In recent experiments, stable and well-behaved NV centers located with depth ranging from 1 to 10 nm from the diamond surface have been observed and the coherence time $T_{2}$ longer than 100 ${\mu}{\text{s}}$ has been achieved for the depth of 5 nm from the diamond surface \cite{apl2,prbr1}. The coherence time of the near-surface NV electron spin can be further increased by high temperature annealing \cite{apl3} and etching away the diamond surface \cite{LovchinskyScience2016}. As shown in Figs. \ref{contour}(c) and \ref{contour}(d), the typical timescale of the fingerprint oscillation of a target spin is much shorter than the NV electron spin coherence time, so the nuclear spin bath would have negligible effect on the positioning. The other NV sensors also have negligible influence on the sensor spin coherence due to the different strain factors of different NV sensors, therefore the spin coherence of the NV sensors matches the contribution solely from the target electron spin [Fig. \ref{three}(b)].

Figure \ref{matching} shows the numerical simulation for positioning a target spin. First, we establish a fingerprint library to store the positions and depths of the dips or peaks in the decoherence patterns caused by a single electron spin located in a large range (5 {\text{nm}}${\leq}R\leq30$ {\text{nm}}, $0\leq\theta\leq\pi/2$). $R$ and $\theta$ are discretized with resolution $dR$=0.02 nm and $d\theta$=0.2$^{\circ}$. Second, we put the diamond tip containing three NV centers close to the target spin and get the respective spin coherence profiles of the three NV centers under CPMG-30 control. Third, by matching the sensor spin coherence oscillations to the fingerprint library, the position of the target spin is restricted to a small spatial region intersected by the small ranges of the three sets of parameters ($R_{i}, \theta_{i}$). The smaller the volume of the intersected region, the higher the resolution in detecting the target spin. In the simulation, the resolution of less than 0.3 nm is achieved.

\begin{figure}
\begin{minipage}[t]{0.5\linewidth}
\centering
\includegraphics[width=1.75in]{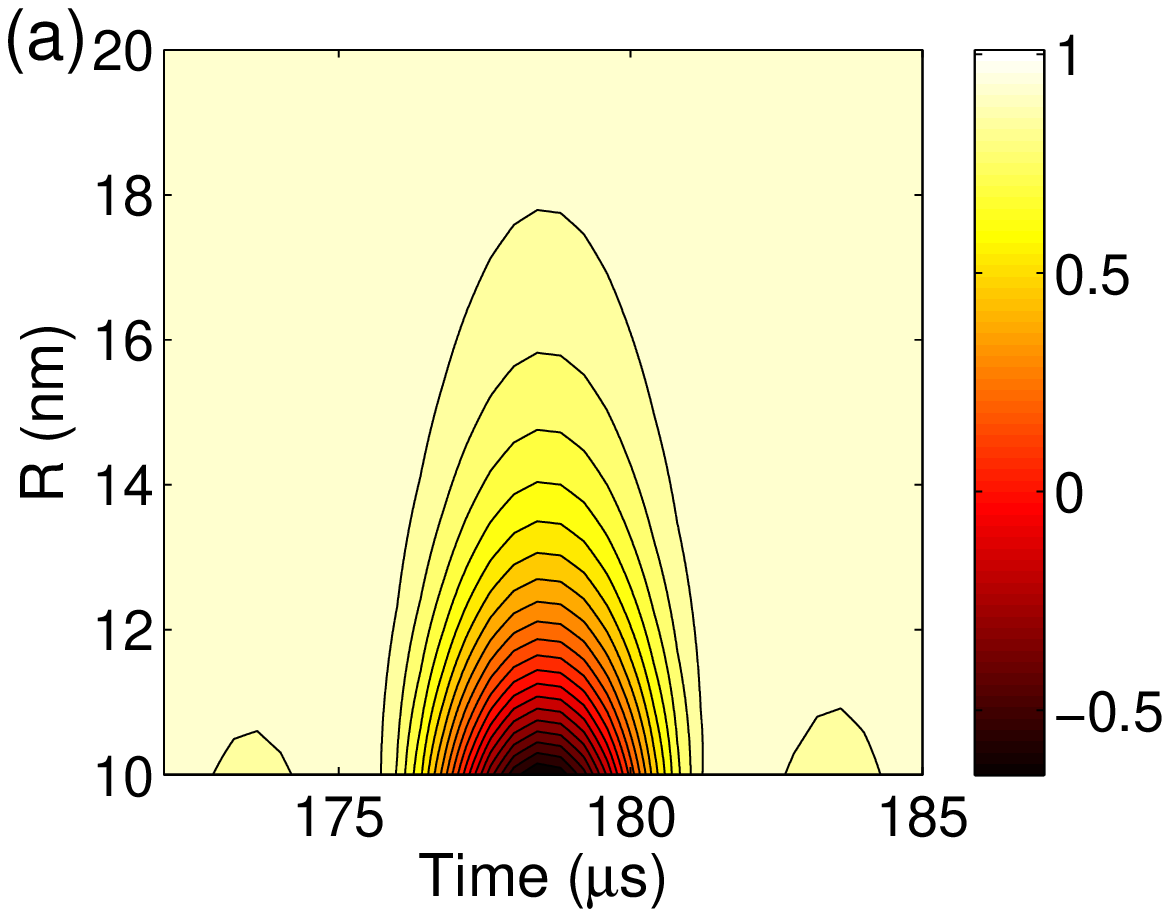}
\end{minipage}%
\begin{minipage}[t]{0.5\linewidth}
\includegraphics[width=1.75in]{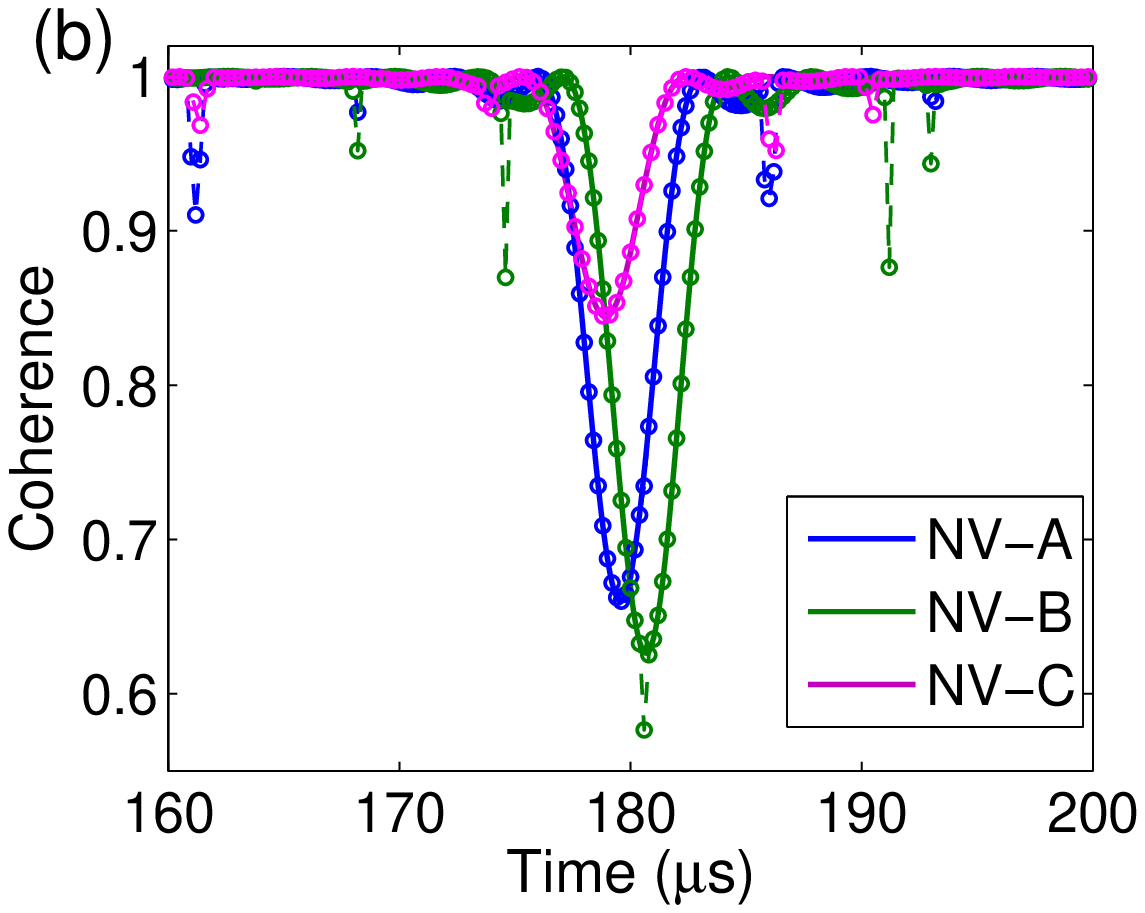}
\end{minipage}
\centering
\caption{Positioning of a single spin about 10$\sim$20 nm away. (a) NV center spin decoherence caused solely by the target spin under CPMG-100 control as a function of time and the distance $R$ with a zenith angle $\theta=30^{\circ}$. (b) Spin coherence of three NV centers under CPMG-100 control (dashed lines with circles) which matches the contribution solely from a single electron spin (solid lines) located at $(R_{A}=13.13$ {\text{nm}}, $ \theta_{A}=23.60^{\circ}, R_{B}=12.57$ {\text{nm}}, $\theta_{B}=13.35^{\circ}, R_{C}=14.73$ {\text{nm}}, $ \theta_{C}=33.88^{\circ})$. Here the additional narrow dips in the detected signal of a specific NV sensor (dashed lines with circles) is caused by the other NV sensors. The parameters are the same as those in Fig. \ref{contour}. }
\label{extend}
\end{figure}

Now we discuss how to extend the detection range of a target spin. The strain-induced transverse anisotropy of the NV sensor leads to the energy splitting of the $|\pm1\rangle_i$ states, and the sensor in the renormalizd eigenstates $|\pm\rangle_i$ have a smaller effective magnetic moment than that of the sensor in states $|\pm1\rangle_i$. Consequently, the decoherence caused by the target spin would be suppressed, which is disadvantageous to the positioning. To increase the detection range, the strain-induced transverse anisotropy should be decreased. Moreover, since the dipolar interaction between the NV sensor spin and the target spin decreases rapidly as the distance between them increases, the magnetic field should be decreased correspondingly to be comparable to the interaction between the sensor spin and target spin. However, the renormalized dipolar interaction decreases almost linearly with the magnetic field. Therefore, the detection range cannot be increased by decreasing the magnetic field. In this case, we can increase the number of CPMG pulses to further amplify the effect of slight difference of the dipolar interaction on the time and depth of the coherence dip. As shown in Fig. \ref{extend}, under CPMG-100 control, the detection range of the multi-spin sensor can be extended to $\sim$20 nm. At such a large distance, we can still achieve sub-nanometer resolution in the simulation. Moreover, with the extension of the detection range, we can choose the near-surface NV centers with larger depth ($>$ 10 nm) so that the surface decoherence effects can be greatly reduced \cite{RomachPRL2015,MyersPRL2014,RosskopfPRL2014}.

To increase the CPMG pulse number with a fixed magnetic field requires a longer coherent evolution time for both the sensor and target spin. For the case in Fig. \ref{extend}, the target spin should have long coherence time greater than $100~\mu$s. However, this requirement can be relaxed by increasing the external magnetic field according to Eq. (\ref{tdip}). A key requirement to realize nanoscale resolution in single spin detection in our scheme is that the external magnetic field should be comparable to the dipolar interaction between the sensor and target spin. If the hyperfine interaction is increased by lowering the strain factors, which may be realized by choosing deeper NV centers \cite{NV-strain} or tuning of the strain factors by electric fields \cite{npb}, then the magnetic field can be increased without decreasing the spatial resolution. For example, if the strain factor $\varepsilon_i$ of the $i$th NV sensor is decreased from 3 MHz to 0.3 MHz, the magnetic field can be increased from 0.1 G to 1 G with the ratio of $\omega_e$ to $\omega_\mathrm{NV} h_i^e/\sqrt{\varepsilon_i^2+\omega_\mathrm{NV}^2}$ unchanged, so the minimal coherent evolution time of the target spin is decreased to about 5 $\mu$s for CPMG-30 and 18 $\mu s$ for CPMG-100.


In Ref. \cite{ar1}, a scanning magnetometer containing one NV center was employed to realize nanoscale magnetic imaging of another target electron spin located about 50 nm away.
There the sensor and target spins were manipulated synchronously by periodic DD control (up to 100 pulses), to constructively accumulate the phase shifts induced on the sensor by the target spin. In our scheme, only the sensor spin is manipulated by DD control, and the sensor spin coherence dip is caused by the intrinsic quantum dynamics of the target spin.

Finally, we discuss some details of the possible experimental realization of our detection scheme. To accurately position a target spin, we have to accurately determine the relative locations, strain parameters and orientations of the multiple NV sensor spins. Recently the depth of shallow NV centers in diamond can be determined with $\sim$ 1 nm resolution by detecting the nuclear magnetic resonance signal of a proton nuclear spin bath placed on the surface \cite{NV-depth}. The strain parameters of the NV sensors has also been determined with $\sim$ 2 kHz accuracy by measuring ODMR spectra of NV sensors at zero or very weak magnetic field \cite{NV-strain}. The orientation of a NV sensor can be determined by measuring the optically detected magnetic resonance spectra (ODMR) in a relatively large magnetic field \cite{NV-orien}, since the sensor electron spin Zeeman splitting would be different for different orientations. Although we assume that all the three NV sensors have the same orientation, the detection scheme can still work for multiple NV sensors of different orientations by just transforming between different local coordinates of the different NV sensors when performing data analysis of different sensor signals. Moreover, the control errors of finite DD pulses can accumulate for a large DD pulse number and degrade the DD performance. In experiments, the XY8-$k$ pulse sequence rather than CPMG pulse sequence can largely suppress the pulse errors \cite{prl4,XY-DD}. Recently the nuclear magnetic resonance of multiple nuclear species has been successfully detected by using a shallow NV sensor under up to XY8-$40$ control (equivalent to CPMG-320 in theory) \cite{nanoNMR}.

\section{Conclusions}
In summary, we have proposed and numerically demonstrated that atomic-scale positioning of single electron spins can be achieved by using multiple NV centers as the sensors. Each NV sensor spin works independently and gathers information about the spatial range of the target spin. By integrating the information provided by multiple NV sensor spins, the position of the target can be accurately determined. The scheme, without requiring spatial scanning or varying the magnetic field direction, may provide an approach to fast and scalable positioning of single spins with sub-nanometer resolution.

\section{acknowledgments}
This work was supported by Hong Kong Research Grants Council - Collaborative CUHK4/CRF/12G and the Chinese University of Hong Kong Vice Chancellor's One-off Discretionary Fund, National Basic Research Program of China (973 Program) under Grant No. G2009CB929300 and National Natural Science Foundation of China under Grant No. 61121491.

\appendix

\section{Diagonalization of the NV center Hamiltonian}
The Hamiltonian of the $i$th NV center in Eq. (\ref{H-NV}) can be written in a matrix form in the basis of the eigenstates $\{|+1\rangle_i ,|0\rangle_i, |-1\rangle_i\}$ as
\begin{equation}
{H}_{\mathrm{NV}_i}=
\left(
  \begin{array}{ccc}
     \Delta+\omega_\mathrm{NV} & 0 & \varepsilon_i   \\
    0 & 0 & 0  \\
    \varepsilon_i & 0 & \Delta-\omega_\mathrm{NV}  \\
  \end{array}
\right),
\label{NV0}
\end{equation}
Diagonalization of this matrix gives the new basis of eigenstates $\{|+\rangle_i, |0\rangle_i, |-\rangle_i\}$ with eigenenergies $\{\Delta+\sqrt{\varepsilon_i^{2}+\omega_\mathrm{NV}^{2}},$ 0, $\Delta-\sqrt{\varepsilon_i^{2}+\omega_\mathrm{NV}^{2}}\}$. The new eigenstates are
\begin{eqnarray}\label{}
   |\pm\rangle_i=\frac{1}{C_{\pm}}\left[\left(\omega_\mathrm{NV}\pm\sqrt{\varepsilon_i^{2}+\omega_\mathrm{NV}^{2}}\right)|+1\rangle_i+\varepsilon_i|-1\rangle_i\right],
\end{eqnarray}
where $C_{\pm}=\sqrt{2\sqrt{\varepsilon_i^{2}+\omega_\mathrm{NV}^{2}}(\sqrt{\varepsilon_i^{2}+\omega_\mathrm{NV}^{2}}\pm\omega_\mathrm{NV})}$.

\section{Derivation of the renormalized pure-dephasing Hamiltonian}
Since the zero-field splitting of the $i$th NV sensor is much larger than its dipolar coupling to the target spin, the other NV sensors and the nuclear spin bath, the state $|0\rangle_i$ can be safely assumed to be decoupled from the states $|\pm1\rangle_i$. Then the effective Hamiltonian in the subspace $\{|+1\rangle_i, |-1\rangle_i\}$ becomes
\begin{equation}
{H}_i=\Delta+
\left(
  \begin{array}{cc}
     \omega_\mathrm{NV}+{h}_{i} &  \varepsilon_i   \\
    \varepsilon_i & -\omega_\mathrm{NV}-{h}_{i}  \\
  \end{array}
\right)+{H}_i^{(0)},
\end{equation}
The diagonal form of this Hamiltonian is obtained as
\begin{eqnarray}\label{}
   {H}_i=\sqrt{\varepsilon_i^{2}+(\omega_\mathrm{NV}+{h}_{i})^{2}}(|+\rangle_i\langle+|-|-\rangle_i\langle-|)+{H}_i^{(0)},
\end{eqnarray}
where we have dropped the zero-field splitting $\Delta$ for simplicity. Since the target spin, NV sensors and nuclear spin bath are generally unpolarized at room temperature, the mean value of the noise field ${h}_{i}$ is much smaller than $\sqrt{\varepsilon_i^{2}+\omega_\mathrm{NV}^{2}}$, so the Hamiltonian can be expanded in the second order of ${h}_{i}$ as
\begin{eqnarray}\label{}
   {H}_i=H_i^{(+)}|+\rangle_i\langle+|+H_i^{(-)}|-\rangle_i\langle-|,
\end{eqnarray}
with
\begin{align}\label{}
   H_{i}^{(\pm)}\approx&H_i^{(0)}\pm\left(\sqrt{\varepsilon_{i}^{2}+\omega_\mathrm{NV}^{2}}+\frac{\omega_e h_i}{\sqrt{\varepsilon_{i}^{2}+\omega_\mathrm{NV}^{2}}}
+\frac{h_i^{2}}{2\sqrt{\varepsilon_{i}^{2}+\omega_\mathrm{NV}^{2}}}\right),
\end{align}
Now we can write the Hamiltonian in the full Hilbert space $\{|+\rangle_i, |0\rangle_i, |-\rangle_i\}$ and get the pure-dephasing Hamiltonian in Eq. (\ref{HR}).

\section{General semiclassical noise model for quantum sensing}
We consider a general Hamiltonian of a spin-1/2 sensor coupled to a target spin cluster \cite{nano3},
\begin{eqnarray}\label{H-noise}
   { H} =   S_z  \beta  + { H_0},
\end{eqnarray}
with
\begin{align}\label{}
  &{ H_0}{\rm{ = }}\sum_{n=1}^{d}{E_n}\left| n \right\rangle \left\langle n \right|, \\
  &  \beta {\rm{ = }}\frac{1}{2}\sum\limits_{m,n} {\left( {{\beta _{mn}}\left| m \right\rangle \left\langle n \right|  +\text{H.c.}} \right)},
\end{align}
where $\beta$ is the noise operator for the sensor, $H_0$ is the free Hamiltonian for the target spin cluster, $d$ is the number of eigenstates of the target spin cluster, $\beta_{mn}=\langle m|\beta|n\rangle$. In the interaction picture set by $H_0$, the time-dependent noise operator is $\beta(t)=e^{iH_0t}\beta e^{-iH_0t}$, and the noise correlation function is just
\begin{eqnarray}\label{}
   C(t)=\langle \beta(t)\beta(0)\rangle=\frac{1}{d}\sum_{m,n}|\beta_{mn}|^{2}e^{i\omega_{mn}t},
\end{eqnarray}
where $\omega_{mn}=E_m-E_n$ and $\langle\cdots\rangle=\mathrm{Tr}[\rho_0\cdots]$ with $\rho_0= d^{ - 1}\sum\nolimits_{m = 1}^d{|m\rangle\langle m|}$. The noise spectrum is the Fourier transform of the noise correlation function \cite{prb3},
\begin{align}\label{}
   C(\omega)=\int_{-\infty}^{\infty}C(t)e^{i\omega t}dt=\frac{2\pi}{d}\sum_{m,n}|\beta_{mn}|^{2}\delta(\omega-\omega_{mn}).
\end{align}
It is easy to demonstrate that $C(t)=C(-t)$ and $C(\omega)=C(-\omega)$. Typically DD control (consisting of a sequence of $\pi$ flips at times $\{t_{1},t_{2}\cdots{t_{N}}\}$ for the sensor evolution from 0 to $t$) is applied to the sensor spin to selectively amplify a specific noise frequency. With the Gaussian noise approximation, the sensor spin decoherence under DD control is \cite{prb3}
\begin{align}\label{L(t)}
   L(t)&=\text{exp}\left[{-\frac{1}{2}\int_{0}^{t}\int_{0}^{t}dt_{1}dt_{2}C(t_{1}-t_{2})f(t_{1})f(t_{2})}\right]  \nonumber \\
       &=\text{exp}\left[{-\frac{1}{2}\int_{-\infty}^{\infty}\frac{d\omega}{2\pi}C(\omega)\frac{F^{2}(\omega,t)}{\omega^{2}}}\right],
\end{align}
where $f(t)=(-1)^{k}$ for $[t_{k},t_{k+1}]$ is the DD modulation function ($t_0$=0, $t_{N+1}=t$), $F(\omega,t)=\omega|\int_{0}^{t}f(t')dt'|^{2}=|\sum_{k=0}^{N}(-1)^{k}(\text{\textit{e}}^{i\omega{t}_{k+1}}-\text{\textit{e}}^{i\omega{t}_{k}})|$ is the DD filter function.

For the $N$-pulse CPMG control, $t_{k}=(2k-1)/2N$ ($k=1,2,\cdots,{N}$) with the pulse delay $2\tau=t/N$ and the filter function is \cite{prb3}
\begin{eqnarray}\label{}
   F(\omega, t){\rm{ = }}\left\{ \begin{array}{l}
 4{{\sin }^2}\left( {\frac{{\omega t}}{{4N}}} \right)\left|\cos \left( {\frac{{\omega t}}{2}} \right){{\cos }^{ - 1}}\left( {\frac{{\omega t}}{{2N}}} \right) \right|,\text{odd}\;N,\\
 4{{\sin }^2}\left( {\frac{{\omega t}}{{4N}}} \right)\left|\sin \left( {\frac{{\omega t}}{2}} \right){{\cos }^{ - 1}}\left( {\frac{{\omega t}}{{2N}}} \right) \right|,\text{even}\;N.
\end{array} \right.
\end{eqnarray}
If the CPMG pulse delay matches one noise frequency $|\omega_{mn}|$, that is $2\tau_{\text{dip}}=(2q-1)\pi/|\omega_{mn}|$ or ${t_{{\rm{dip}}}} = \pi (2q - 1)N/|\omega_{mn}|$ ($q=1,2,\cdots$), the noise with this frequency is amplified by CPMG control and the sensor spin coherence shows sharp dips. Since the filter function at the coherence dips is $F({\omega_{mn}},{t_{{\rm{dip}}}}) = 2N$, the sensor coherence dip depth corresponding to $|\omega_{mn}|$ as a function of CPMG pulse number is
\begin{align}\label{}
   L^i_{\text{dip}}(N)\approx\text{exp}\left(-\frac{4N^{2}|\beta_{mn}|^{2}}{d\omega^2_{mn}}\right).
\end{align}
The semiclassical noise model is valid for the weak sensor-target coupling regime ($|\beta_{mn}|\ll |\omega_{mn}|$) and relatively small DD pulse number. Moreover, the sensor coherence dip depth from the semiclassical noise model is always positive while the one from exact quantum model can be negative \cite{nano2}.

\section{Sensing a single target electron spin by the NV sensor spins}
The Hamiltonian for the $i$th NV sensor coupled to a remote target electron spin is
\begin{align}\label{}
   H&=\left(\lambda_i h_i^{e}+H_{\text{tar}}\right)|+\rangle_i\langle+|+H_{\text{tar}}|0\rangle_i\langle0| \nonumber \\
    &=\frac{\lambda_i h_i^{e}}{2}\sigma_z+\frac{\lambda_i h_i^{e}}{2}+H_{\text{tar}},
\end{align}
where $\sigma_i^z=|+\rangle_i\langle+|-|0\rangle_i\langle0|$ is the pauli operator in the subspace $\{|0\rangle_i, |+\rangle_i\}$, $\lambda_i=\omega_\mathrm{NV}/\sqrt{\varepsilon^{2}_{i}+\omega_\mathrm{NV}^{2}}$ is the renormalization factor, $h_i^{e}=(\hat{\mathbf{z}}\cdot{\mathbb{A}_{i,e}})\cdot{\mathbf{S}_{e}}={\mathbf{A}_{i,e}}\cdot{\mathbf{S}_{e}}$ and $H_{\text{tar}}=\omega_e S_e^z$.
In comparison with the general Hamiltonian in Eq. (\ref{H-noise}), the noise operator for the $i$th NV center is
\begin{align}\label{noise}
 \beta=\lambda_i h_i^{e}=\lambda_i{\mathbf{A}_{i,e}}\cdot{\mathbf{S}_{e}},
\end{align}
and the free Hamiltonian of the target spin is
\begin{align}\label{free}
H_0=\lambda_i h_i^{e}/2+H_{\text{tar}}=\frac{\lambda_i}{2}{\mathbf{A}_{i,e}}\cdot{\mathbf{S}_{e}}+\omega_e S_e^z.
\end{align}

\subsection{Semiclassical noise model}
In the weak-coupling regime ($\|\beta\|\ll \|H_0\|$ or $|\mathbf{A}_{i,e}|\ll \omega_e$), the noise spectrum of the target electron spin can be simplified as
 \begin{eqnarray}\label{}
   C(\omega)\approx\frac{\pi}{4}(\lambda_iA_{i,e}^\perp)^{2}[\delta(\omega-\omega'_{e})+\delta(\omega+\omega'_{e})],
\end{eqnarray}
where $\omega'_{e}\approx\lambda_iA_{i,e}^z/2+\omega_{e}$ and $A_{i,e}^\perp=\sqrt{(A_{i,e}^x)^2+(A_{i,e}^y)^2}$ (the zero-frequency noise has no contribution to the sensor decoherence under DD control and is neglected here). We can see that the noise frequency of the target spin is the Zeeman frequency plus a shift caused by the hyperfine interaction with the NV sensor spin. So the sensor coherence dip occurs at times ${t_{{\rm{dip}}}} = \pi (2q - 1)N/\omega'_e$, and the coherence dip depth as a function of the CPMG pulse number is
\begin{align}\label{Hdip-c}
   L^i_{\text{dip}}(N)&\approx\text{exp}\left(-\frac{N^{2}(\lambda_iA_{i,e}^\perp)^{2}}{2(\omega'_e)^{2}}\right).
\end{align}

\subsection{Quantum decoherence model}
In the quantum decoherence model, the NV sensor spin decoherence is caused by the bifurcated evolution of the target spin conditioned on the sensor spin state \cite{CE,prb4,pair}. So the decoherence of the $i$th NV sensor spin under DD control is
\begin{align}\label{H-tar}
   {L}_i{\rm{(}}t{\rm{) = }}\frac{1}{{{2}}}{\rm{Tr}}\left[ {{{\left( { U^{(0)}_N} \right)}^\dag } U^{(+)}_N} \right],
\end{align}
with
\begin{subequations}
\begin{align}\label{}
   {U}^{(+)}_{N}(t)=\cdots e^{-i(H_0-\beta/2)(t_2-t_1)}e^{-i(H_0+\beta/2)t_{1}}, \\
   {U}^{(0)}_{N}(t)=\cdots e^{-i(H_0+\beta/2)(t_2-t_1)}e^{-i(H_0-\beta/2)t_{1}}.
\end{align}
\end{subequations}
Eq. (\ref{H-tar}) can be solved exactly for any DD control, and in the weak coupling regime ($\|\beta\|\ll \|H_0\|$) the sensor coherence dip depth can be written in a general form for a spin-1/2 sensor \cite{dip}. In the following part, we adapt the derivation in Ref. \cite{dip} to get an expression for the NV sensor coherence dip depth as a function of the CPMG pulse number.

Since $\|\beta\|\ll \|H_0\|$, we adopt the interaction picture, in which the noise operator is defined as
\begin{align}\label{}
     \beta (t) &= {e^{i{{ H}_0}t}} \beta {e^{ - i{{ H}_0}t}} \nonumber \\
    &\approx {\frac{\lambda_i A_{i,e}^{\perp}}{2}\left[ { S_e^+ {e^{i({\omega'_e}t-\alpha)}} + S_e^- {e^{ - i({\omega'_e}t-\alpha)}}} \right]}+\lambda_i A_{i,e}^z  S_e^z,
\end{align}
where $\omega'_{e}$ and $A_{i,e}^\perp$ are the same as those defined in the last part, $S_e^{\pm}=S_e^x\pm iS_e^y$, and $\alpha=\arctan(A_{i,e}^y/A_{i,e}^x)$. Here we have made an approximation $H_0\approx (\lambda_iA_{i,e}^z/2+\omega_{e})S_e^z$.
Then ${U}^{(+)}_{N}$ and ${U}^{(0)}_{N}$ can be written as
\begin{subequations}
\begin{align}\label{}
   {U}^{(+)}_{N}(t)={e^{ - i{{ H}_0}t}} T{e^{- \frac{i}{2}\int_0^t {f(t')}  \beta (t')dt'}}, \\
   {U}^{(0)}_{N}(t)={e^{ - i{{ H}_0}t}} T{e^{+ \frac{i}{2}\int_0^t {f(t')}  \beta (t')dt'}},
\end{align}
\end{subequations}
where $f(t)=(-1)^{k}$ for $[t_{k},t_{k+1}]$ is the DD modulation function and ${T}$ is the time-ordering operator.

Now we try to simplify ${U}^{(+)}_{N}$ and ${U}^{(0)}_{N}$. According to the Magnus expansion \cite{magnus-1}, a general time-evolution operator can be expanded as
\begin{eqnarray}\label{}
    U(t) =  T{e^{ - i\int_0^t { H(t')dt'} }} = \exp \left( {\sum\limits_{l = 1}^\infty  {{{ \Omega }_l}(t)} } \right),
\end{eqnarray}
with the first-order and second-order Magnus terms
\begin{align}\label{}
   &{ \Omega _1}(t) =  - i\int_0^t { H(t')dt'}, \\
   &{ \Omega _2}(t) =  - \frac{1}{2}\int_0^t {d{t_1}\int_0^{{t_1}} {d{t_2}} \left[ { H({t_1}),\; H({t_2})} \right]}.
\end{align}
For our specific model, the first-order Magnus term is
\begin{align}\label{mag-1}
   { \Omega _1}(t) &=  - \frac{i\lambda_iA_{i,e}^{\perp}}{4}\int_0^t {f(t')\left[ { S_e^+ {e^{i{(\omega'_e}t'-\alpha)}} +  S_e^- {e^{ - i{(\omega'_e}t'-\alpha)}}} \right]dt'}  \nonumber \\
   & =  - {\frac{{i\lambda_i{A_{i,e}^{\perp}}}}{{2{\omega'_e}}}} F({\omega'_e},t) S_e^{(\xi-\alpha)},
\end{align}
with
\begin{subequations}\label{}
\begin{align}\label{}
   & F(\omega'_e,t) = \omega'_e \left|\int_0^t {f(t'){e^{i\omega'_e t'}}dt'}\right|, \\
   & {e^{i\xi }} = \left(\omega'_e \int_0^t {f(t'){e^{i\omega'_e t'}}dt'}\right)/F(\omega'_e,t), \\
   &  S_e^{\xi}  = \cos \xi S_e^x - \sin \xi S_e^y,
\end{align}
\end{subequations}
Note that the term $\lambda_iA_{i,e}^z  S_e^z$ is averaged out by the DD control, so it vanishes in Eq. (\ref{mag-1}). In the weak-coupling regime, it has been shown that ${ \Omega _2}(t)\ll{ \Omega _1}(t)$ \cite{dip,magnus-1}. So we can use the first-order Magnus expansion to approximate ${U}^{(+)}_{N}$ and ${U}^{(0)}_{N}$ as
\begin{subequations}
\begin{align}\label{}
   &{U}^{(+)}_{N}(t)\approx \exp \left( { - i{{ H}_0}t} \right)\exp \left( { - i{\frac{{\lambda_i{A_{i,e}^{\perp}}}}{{2{\omega'_e}}}} F({\omega'_e},t) S_e^{(\xi-\alpha)} } \right), \\
   &{U}^{(0)}_{N}(t)\approx \exp \left( { - i{{ H}_0}t} \right)\exp \left( { + i{\frac{{\lambda_i{A_{i,e}^{\perp}}}}{{2{\omega'_e}}}} F({\omega'_e},t) S_e^{(\xi-\alpha)} } \right),
\end{align}
\end{subequations}
So the qubit decoherence is
\begin{align}\label{}
   {L}_i{\rm{(}}t{\rm{) = }}\frac{1}{{{2}}}{\rm{Tr}}\left[ {{{\left( { U^{(0)}_N} \right)}^\dag } U^{(+)}_N} \right] \approx \cos {\left[\frac{{\lambda_i{A_{i,e}^{\perp}}}}{{2{\omega'_e}}}F({\omega' _e},t)\right]}.
\end{align}
At the sensor coherence dips ${t_{{\rm{dip}}}} = \pi (2q - 1)N/\omega'_e$ for the CPMG control, $F({\omega' _e},t)=2N$, so the sensor coherence dip depth as a function of the CPMG pulse number is
 \begin{align}\label{Hdip-q}
 L^i_{\text{dip}}(N){\approx}\cos\left(\frac{\lambda_iA_{i,e}^{\bot}N}{\omega'_e}\right)=\cos\left(\frac{\lambda_iA_{i,e}^{\bot}N}{\omega_e+\lambda_iA_{i,e}^{z}/2}\right).
 \end{align}
For a small CPMG pulse number, Eq. (\ref{Hdip-q}) from the quantum model agrees with Eq. (\ref{Hdip-c}) as derived in the semiclassical noise model. But for a large CPMG pulse number, the coherence dips in Eq. (\ref{Hdip-q}) can be negative, while Eq. (\ref{Hdip-c}) always predicts positive coherence dips. Eq. (\ref{Hdip-q}) agrees almost exactly with the exact quantum results \cite{dip}.


\end{document}